\documentclass[twocolumn,showpacs,preprintnumbers,amsmath,amssymb,APSl,prd,nofootinbib,superscriptaddress]{revtex4-1}

\usepackage{bm}
\usepackage{mathrsfs}
\usepackage{xcolor,color,graphicx,graphics}
\usepackage[all]{xy}
\usepackage{epsfig,subfigure}
\usepackage{latexsym,amssymb,amsmath,amsfonts} 
\usepackage[english]{babel} 
\usepackage[OT1]{fontenc}
\usepackage[latin1]{inputenc}
\usepackage{makeidx}
\usepackage{hyperref}
\usepackage{color,graphicx,graphics,wrapfig,epsf}
\usepackage{float}
\restylefloat{table}
\usepackage[colorinlistoftodos]{todonotes}
\usepackage{blindtext, xcolor}
\usepackage{verbatim}
\definecolor{red}{rgb}{1,0,0}

\def\+{^\dagger}

\def\<{\leftarrow}
\def\>{\rightarrow}

\def\({\left(}
\def\){\right)}

\def\ca{\mathcal{A}}

\def\cb{\mathcal{B}}
\def\cv{\mathcal{V}}

\def\hr{\hat{r}}

\def\hi{\mathcal{I}}

\def\bb{\bibitem}

\def\bg{\bar{g}}
\def\bp{\bar{\Phi}}
\def\baa{\bar{\mathcal{A}}}
\def\bb{\bar{\mathcal{B}}}
\def\bv{\bar{\mathcal{V}}}
\def\br{\bar{r}}

\newcommand{\bi}{\begin{itemize}} 				\newcommand{\ei}{\end{itemize}}
\newcommand{\benu}{\begin{enumerate}} 		\newcommand{\enu}{\end{enumerate}}
\newcommand{\bd}{\begin{dinglist}{0}}     \newcommand{\ed}{\end{dinglist}}
\newcommand{\bfig}{\begin{figure}[htbp]}  \newcommand{\efig}{\end{figure}}
        			
\newcommand{\bc}{\begin{center}} 				  \newcommand{\ec}{\end{center}}
\newcommand{\be}{\begin{equation}} 				\newcommand{\ee}{\end{equation}}
\newcommand{\bsub}{\begin{subequations}}  \newcommand{\esub}{\end{subequations}}
\newcommand{\ben}{\begin{eqnarray}} 			\newcommand{\een}{\end{eqnarray}}
\newcommand{\ba}[1]{\begin{array}{#1}} 		\newcommand{\ea}{\end{array}}
\newcommand{\bea}{\begin{equation}\begin{array}{rcl}}
\newcommand{\eea}{\end{array}\end{equation}}

\begin{document}
\title{Invariant quantities of Scalar-Tensor Theories for stellar structure}

\author{Aleksander Kozak}
\email{email}
\affiliation{Institute of Theoretical physics, University of Wroclaw, pl. Maxa Borna 9, 50-206 Wroclaw, Poland
}
\author{Aneta Wojnar}
\email{aneta.magdalena.wojnar@ut.ee}
\affiliation{Laboratory of Theoretical Physics, Institute of Physics, University of Tartu,
W. Ostwaldi 1, 50411 Tartu, Estonia
}

\begin{abstract}
We present the relativistic hydrostatic equilibrium equations for a wide class of gravitational theories possessing a scalar-tensor representation. It turns out that the stellar structure equations can be written with respect to the scalar-tensor invariants, allowing to interpret their physical role.
\end{abstract}

\maketitle
\tableofcontents

\section{Introduction}

Scalar-Tensor Theories (STT) are a class of so-called Extended Theories of Gravity, whose main aim is to go beyond standard theory of gravity formulated by Einstein in order to account for certain phenomena that cannot be satisfactorily explained by General Relativity (GR) itself \cite{noj1}-\cite{fuj}. They introduce an additional mediator of gravitational interaction, a scalar field, coupled non-minimally to the curvature and, possibly, matter fields. Historically, this class of theories was considered first by Jordan \cite{jordan} and then by Brans and Dicke \cite{dicke}, who added a scalar field in order to incorporate Mach's Principle into the theory; the massless field, sourced by matter distribution (trace of energy-momentum tensor), acted as an effective gravitational constant. The first attempts to include a new field were, however, lacking theoretical motivation; only then it was shown that such a field, coupled non-minimally to the curvature, arises  naturally in the low-energy limit of theories considered fundamental, such as string theory \cite{capo}, for instance.

STT found applications in various contexts. For example, it can be demonstrated that $f(R)$ theories of gravity \cite{capo2011}-\cite{nojiri2003}, where one replaces the Einstein-Hilbert Lagrangian with a function of the curvature, have an equivalent scalar-tensor (ST) representation. A noteworthy example of $f(R)$ gravity is the Starobinsky model \cite{staro1980}, in which one adds a quadratic correction in order to explain inflationary behavior of the early universe. This model corresponds strictly to a ST theory with a specific potential. Other examples include attempts at accounting for accelerated expansion of the universe, since the field can act as quintessence \cite{bartolo1999}. 

 The theories are analyzed both in the metric, and in the so-called Palatini formalism (or metric-affine), where one decouples metric structure from the affine structure of space-time. Recently, the Palatini approach to ST gravity has been gaining more attention \cite{Ulf}-\cite{vecthorn}. Palatini STT theories were analyzed in the context of inflation \cite{racc}-\cite{stach}. Also, a recent paper proves that one can construct an $f(R)$ theory \`{a} la Palatini which evades Ostrogradsky's instability, is renormalizable and invariant under conformal change of the metric in its high energy limit and reduces to GR for small energies \cite{coumbe}. It was also shown that for certain models, both approaches lead to the same predictions regarding certain observables related to inflation, such as tensor-to-scalar ratio and scalar spectral index \cite{jarv2020}.

When analyzing STT, one usually makes an extensive use of Weyl (or conformal) transformations of the metric tensor \cite{flanagan2004}. Conformal transformations of the metric tensor establish a mathematical equivalence between so-called conformal frames, but usually the description of physics is frame-dependent, since Weyl transformations are not symmetries of Nature. Therefore, in case of STT, it is important to establish which frame is considered physical. The two most commonly used in the literature frames are 'Einstein' and 'Jordan' \cite{kuusk22015}-\cite{kamen2015}; the main difference between them is the nature of coupling of the scalar field. In the Einstein frame, the field is coupled to matter part of the action, whereas in the Jordan frame, the field is coupled to the curvature. 

It is possible, however, to define certain quantities which preserve their form under conformal transformations \cite{tartu}-\cite{karam2018}. Such quantities, introduced in \cite{tartu}, consist of functions of the scalar field defining a conformal frame. One can also construct an invariant metric and, in the Palatini case, an invariant connection, allowing different conformal observers to measure distances and calculate geodesics in the same way \cite{akab}. The advantage of introducing such quantities is also of different type: they allow to classify mathematically equivalent theories, since a class of conformally-related frames will yield the same values of invariants.

Stars in the scalar-tensor theories framework, relativistic \cite{hor}-\cite{dam} and non-relativistic ones \cite{cl}-\cite{ros}, were widely studied in literature; however, the invariant theory has not been applied yet to stellar objects. In this paper, we aim at using the results of \cite{tartu} in order to analyze stellar structure in ST theories in a way independent of the choice of conformal metric. This will be achieved by expressing all relevant equations describing structure of stars in terms of invariants. Such a procedure will allow one not only to easily compare different conformal frames, but also inscribes in the way of reasoning concerning the issue of physicality of different frames described above.

 The paper is organized as follows: First, we start by writing the action for STT in the most general parametrization and then, upon introducing transformation formulae for the metric, scalar fields and functions thereof, we write out most commonly used invariants. Having those, we obtain the Tolman-Oppenheimer-Volkoff (TOV) equation in the invariant Einstein frame, in which the calculations are particularly simple. Then, we transform to the Jordan frame and obtain relation between the pressure and other quantities entering the theory. That should allow us to interpret the extra terms written with respect to the invariants appearing in the modified/extended TOV equation.

 For the reader's convenience, we present physical quantities, parameters and coordinates in different frames in the table below:

\begin{table}
\caption{Summary of quantities used in different frames.}
\centering
\begin{tabular}{ll }
\hline\noalign{\smallskip}
\text{Frame} & \text{Indication} \\
\noalign{\smallskip}\hline\noalign{\smallskip}
\text{General Wagoner parametrization} & $\text{bar}$  \\

\text{Jordan (invariant)} & none  \\

\text{Einstein (invariant)} & $\text{hat}$  \\
\noalign{\smallskip}\hline
\end{tabular}
\end{table}

\section{Invariant quantities}

\subsection{General frame}
The action for the class of scalar-tensor theories of gravity written in the Wagoner parametrization takes the following form:

\begin{equation}\label{scalarTensorAction}
\begin{split}
S[\bar{g}_{\mu\nu}, \bp, \chi]& =  \frac{1}{2\kappa^2}\int_\Omega d^4x \sqrt{-\bg} \Big[\baa(\bp)\bar{R}- \bb(\bp)\bg^{\mu\nu}\partial_\mu\bp\partial_\nu\bp\\
&- \bv(\bp) \Big]+ S_\text{matter}\left[e^{2\bar{\alpha}(\bp)}\bg_{\mu\nu}, \chi\right],
\end{split}
\end{equation}
where $\{ \baa,\bb,\bv,\bar{\alpha} \}$ are four arbitrary functions of the scalar field, providing, together with the choice of the metric and the scalar field the so-called 'conformal frame'. The function $\baa$ describes the coupling between the scalar field and the curvature, $\bb$ is the kinetic coupling, $\bv$ - the self-interaction potential of the scalar field, and $\bar{\alpha}$ - an anomalous coupling between matter and the scalar field. The constant $\kappa^2=8\pi Gc^{-4}$.

\noindent The equations of motion derived from this action are the following \cite{tartu}:

\begin{subequations}
\begin{align}
\begin{split}
& \baa \bar{G}_{\mu\nu} + \left(\frac{1}{2}\bb  + \baa''\right)\bg_{\mu\nu}\bg^{\alpha\beta}\partial_\alpha\bp\partial_\beta\bp \\
& -\left(\bb + \baa''\right)\partial_\mu\bp\partial_\nu\bp - \baa'(\bg_{\mu\nu}\bar{\Box} - \bar{\nabla}_\mu\bar{\nabla}_\nu)\bp \\
&\qquad +\frac{1}{2}\bv \bg_{\mu\nu} = \kappa^2 \bar{T}_{\mu\nu},
\end{split} \\
\begin{split}
& 2[3(\baa')^2 +2\baa\bb ]\bar{\Box}\bp + \frac{d [3(\baa')^2 +2\baa\bb]}{d\Phi} (\partial\bp)^2\\
&+4\baa'\bv-2\baa\bv'=2\kappa^2\,\bar{T} (\baa'-2\bar{\alpha}' \baa)\,.
\end{split}\label{scFiEOM}
\end{align}
\end{subequations}
The action (\ref{scalarTensorAction}) is form-invariant under the conformal change of the metric tensor, accompanied by a re-definition of the scalar field:
\begin{center}
 \begin{equation}
 \bg_{\mu\nu} = e^{2\bar{\bar{\gamma}}(\bar{\bp})}\bar{\bg}_{\mu\nu},\quad \bp = \bar{\bar{f}}(\bar{\bp})
\end{equation}   
\end{center}

if the four functions of the scalar field transform in the following way:

\begin{subequations} \label{eqns7}
				\begin{align}
				& \bar{\bar{\mathcal{A}}}(\bar{\bar{\Phi}})=e^{2\bar{\bar{\gamma}}(\bar{\bp})}\baa(\bar{\bar{f}}(\bar{\bp})), \\
				\begin{split}
				& \bar{\bar{\mathcal{B}}}(\bar{\bp})=e^{2\bar{\bar{\gamma}}(\bar{\bp})}\Bigg(\Big(\frac{d\bp}{d\bar{\bp}}\Big)^2\bb(\bar{\bar{f}}(\bar{\bp}))  -6\Big(\frac{d\bar{\bar{\gamma}}}{d\bar{\bp}}\Big)^2\baa(\bar{\bar{f}}(\bar{\bp}))\\
				&-6\frac{d\bar{\bar{\gamma}}}{d\bar{\bp}}\frac{d\baa}{d\bp}\frac{d\bp}{d\bar{\bp}}\Bigg),
				\end{split} \\
				& \bar{\bar{\mathcal{V}}}(\bar{\bp})=e^{4\bar{\bar{\gamma}}(\bar{\bp})}\bv(\bar{\bar{f}}(\bar{\bp})),\\
				& \bar{\bar{\alpha}}(\bar{\bp})=\bar{\alpha}(\bar{\bar{f}}(\bar{\bp}))+\bar{\bar{\gamma}}(\bar{\bp}).
				\end{align}
\end{subequations}

It is possible to construct out of these functions quantities which have the same functional dependence in every conformal frame. Such quantities are called 'conformal invariants' \cite{tartu}, and their usefulness consists in possibility of expressing certain quantities in a frame-independent way. All mathematically equivalent frames (i.e. relatable by a conformal transformation and a re-definition of a scalar field) will share the same invariants, so it is possible to label classes of equivalent theories by using them. The invariants which will be used in this paper are given below\footnote{Please notice that the invariants are defined differently in \cite{tartu}, where $\mathcal{I}_1 = \frac{e^{2\alpha}}{\ca}$, and the invariant $\mathcal{I}$ playing role of scalar field in the Einstein frame is denoted $\mathcal{I}_3$.}:

\begin{subequations}
\begin{align}
& \mathcal{I}_1 = \frac{\baa}{e^{2\bar{\alpha}}}, \\
& \mathcal{I}_2 = \frac{\bv}{\baa^2},\label{inv2} \\
& \frac{d\tilde{\mathcal{I}}}{d\bp} = e^{-\bar{\alpha}}\sqrt{\pm\left(\bb-6\left(\frac{d\bar{\alpha}}{d\bp}\right)^2\baa + 6\frac{d\bar{\alpha}}{d\bp}\frac{d\baa}{d\bp}\right)},\\
& \frac{d\mathcal{I}}{d\bp}= \sqrt{\pm\frac{2\baa\bb+3(\baa')^2}{2\baa^2}}.
\end{align}
\end{subequations}
The first invariant tells us if there is a non-minimal coupling present in the theory. The second invariant generalizes the notion of potential of the scalar field, while the third one can play a role of an invariant scalar field in the Jordan frame, and becomes zero for metric $f(R)$ theory (but it does not necessarily mean that the field has no dynamics). The last one can be treated as a scalar field in the Einstein frame; if it vanishes, then the scalar field is non-dynamical (as it happens in case of Palatini $f(R)$ theories, where $\mathcal{I} = 0$). In our convention, we do not write any subscript in $\mathcal{I}$ for convenience and to indicate its importance. The plus/minus sign corresponds to positive and negative value of the expression under the square root.

In what follows, we will examine a static, spherical-symmetric objects whose metric is given as usually:
\begin{equation}\label{gen_metric}
\bar{g}_{\mu\nu} = \text{diag}\left(-  \bar{b}(\bar{r}),  \bar{a}(\bar{r}),  \bar{r}^2, \bar{r}^2 \sin{\bar{\theta}}\right).
\end{equation}
Since all quantities are assumed to be dependent on the radial coordinate only, the conformal transformation preserves the static, spherical symmetric geometry. We will always write down the coordinate transformations in the further parts of the paper.


\subsection{Einstein frame}

In order to make use of the invariant quantities introduced earlier in this paper, let us choose the following invariant metric:
\begin{equation}\label{choice1}
\hat{g}_{\mu\nu} = \baa \bg_{\mu\nu}. 
\end{equation}
Written in components, the metric $\hat{g}$ reads:
\begin{equation}
\hat{g}_{\mu\nu} = \text{diag}\left(- \baa \bar{b}(\bar{r}), \baa \bar{a}(\bar{r}), \baa \bar{r}^2, \baa \bar{r}^2 \sin{\bar{\theta}}\right).
\end{equation}
We may want to introduce a new radial coordinate and re-define the ${a, b}$ functions in order to preserve the form (\ref{gen_metric}) of the metric. Such a change of variables is defined as follows (remembering that the scalar field is the function of the radial component only):
\begin{equation}
\hat{r} = \sqrt{\baa}\:\bar{r} \quad \rightarrow \quad d\bar{r} = \sqrt{\frac{1}{\baa}}\left(-\frac{\hat{r}}{2}\partial_{\hr} \ln \baa + 1\right)d\hat{r}.
\end{equation}
Let us notice that the new radial component is itself an invariant. The new metric will take the form:
\begin{equation}
\hat{g}_{\mu\nu} = \text{diag}\left(- \hat{b}(\hat{r}), \hat{a}(\hat{r}), \hat{r}^2, \hat{r}^2 \sin{\theta}\right),
\end{equation}
where the new metric components $\hat a$ and $\hat b$ are related to the general ones as: 
\begin{subequations}
\begin{align}
& \hat{b} = \baa\: \bar{b}, \\
& \hat{a} = \left(-\frac{\hat{r}}{2}\partial_{\hr} \ln \baa + 1\right)^2 \bar{a}.
\end{align}
\end{subequations}
Let us notice that the following quantity:
\begin{equation}
    \frac{1}{\hat{a}}\left(\frac{d}{d\hr}\right)^2 = \frac{1}{\baa\bar{a}}\left(\frac{d}{d\bar{r}}\right)^2
\end{equation}
is also an invariant. Therefore, if any equation contains (functions of) invariant quantities and the derivative with respect to the radial component precisely in the form shown above, then it must be also conformally invariant.

The choice (\ref{choice1}) of the invariant metric results in the following action for the theory:
\begin{equation}\label{invariantAction}
\begin{split}
S[\hat{g}_{\mu\nu}, \mathcal{I}, \chi]& =  \frac{1}{2\kappa^2}\int_\Omega d^4x \sqrt{-\hat{g}}\left[\hat{R} - \hat{g}^{\mu\nu}\partial_\mu\mathcal{I}\partial_\nu\mathcal{I} - \mathcal{I}_2\right] \\
&+ S_\text{matter}\left[\frac{1}{\mathcal{I}_1}\hat{g}_{\mu\nu}, \chi\right].
\end{split}
\end{equation}
In this frame, the invariant $\mathcal{I}$ plays the role of the scalar field, and all invariants are now functions of it.

As we can see, in the invariant Einstein frame, the action takes a particularly simple form. The coupling between the curvature and the scalar field is now non-existent, and the kinetic coupling has a constant value. Such a simplicity is achieved at a price: there is an anomalous coupling between the scalar field and matter part of the action, which may lead to undesirable consequences, such as violation of the Weak Equivalence Principle. Nevertheless, it is convenient to perform the calculations in the Einstein frame and then transform the result back to the frame which is deemed 'physical'. 

The field equations resulting from this action can be obtained quickly by comparing the action (\ref{invariantAction}) with the action (\ref{scalarTensorAction}) and identifying the corresponding scalar field functions in the field equations. The result is the following:

\begin{subequations}\label{invEinEOM}
\begin{align}
 \hat{G}_{\mu\nu} +&\frac{1}{2}\hat{g}_{\mu\nu}\hat{g}^{\alpha\beta}\partial_\alpha\mathcal{I}\partial_\beta\mathcal{I} - \partial_\mu\mathcal{I}\partial_\nu\mathcal{I} + \frac{1}{2}\hat{g}_{\mu\nu}\mathcal{I}_2 = \kappa^2 \hat{T}_{\mu\nu}, \\
& \hat{\Box}\mathcal{I} - \frac{1}{2}\frac{d\mathcal{I}_2}{d\mathcal{I}} = \kappa^2\frac{1}{\mathcal{I}_1}\frac{d\mathcal{I}_1}{d\mathcal{I}}\hat{T},\label{invEinEOM2}
\end{align}
\end{subequations}

\noindent where $\hat{T}=\hat{g}^{\mu\nu}\hat{T}_{\mu\nu} = -\hat{\rho} + 3\hat{p} \equiv \frac{1}{\mathcal{\bar{A}}^2}(\bar{\rho} + 3\bar{p})$.

Information about different ST theories, characterized by a particular choice of the functions $\{\baa,\bb,\bv,\bar{\alpha}\}$, is stored in two invariants $\{\mathcal{I}_1, \mathcal{I}_2\}$ being now a function of the third invariant, $\mathcal{I}$. Therefore, the knowledge of the functions $\mathcal{I}_1(\mathcal{I})$ and $\mathcal{I}_2(\mathcal{I})$ allows one to reconstruct the original theory after two of the four functions of the field have been fixed. 

The energy-momentum tensor is not conserved\footnote{See the review \cite{thiago} on non-conservative theories of gravity.} in this frame:
\begin{equation}
\hat{\nabla}^\mu\hat{T}_{\mu\nu} = \frac{1}{2}\partial_\nu(\ln \mathcal{I}_1)\hat{T}.
\end{equation}
The relation between the trace of the energy-momentum tensor in different frames are given in (\ref{conftrans}) and (\ref{conftrans2}).

\section{Stellar structure equations}
In the following section we will derive the Tolman-Oppenheimer-Volkoff equation written with respect to the invariants of the scalar-tensor theories. Considering the Newtonian limit of TOV equations, we will also provide the non-relativistic hydrostatic equilibrium equation.

In order to do so, let us notice that the field equations (\ref{invEinEOM}) have a particular form discussed in \cite{mim}-\cite{mim3}, that is:
\begin{equation}\label{mod1}
 \bar{\sigma}(\bar{\Psi}^i)(\bar{G}_{\mu\nu}-\bar{W}_{\mu\nu})=\kappa^2 \bar{T}_{\mu\nu},
\end{equation}
where $\bar{G}_{\mu\nu}$ represents the Einstein tensor while the factor $\bar{\sigma}(\bar{\Psi}^i)$ is a coupling to the gravity, $\bar{\Psi}$ can stand for, for instance, curvature invariants or other fields, like scalar ones in our case. The tensor $\bar{W}_{\mu\nu}$ is a symmetric tensor which can be treated as an additional geometrical term, whose form depends on theory of gravity one is interested in. The equations (\ref{mod1})
do not take into account other field equations which can be obtained by varying a specific Lagrangian with respect to, for example, scalar fields, or an independent connection.

 For such a representation, the TOV equations can be written in a schematic way as: \cite{an_he}\footnote{Let us notice that the authors of \cite{an_he} use a different $\kappa$ convention in which $\kappa$ is negative \cite{weinberg}.}:
\begin{align}\label{tov}
  \left(\frac{\bar{\Pi}}{\bar{\sigma}}\right)'=&-\frac{G\mathcal{M}}{c^2\bar{r}^2}\left(\frac{c^2\bar{Q}}{\bar{\sigma}}+\frac{\bar{\Pi}}{\bar{\sigma}}\right)
  \left(1+\frac{4\pi \bar{r}^3\frac{\bar{\Pi}}{\bar{\sigma}}}{c^2\mathcal{M}}\right)\bar{a}(\bar{r})\nonumber\\
  & -\frac{2\bar{\sigma}}{\kappa^2 \bar{r}}\left(\frac{\bar{W}_{\theta\theta}}{\bar{r}^2}-\frac{\bar{W}_{\bar{r}\bar{r}}}{\bar{a}}\right)
\end{align}
with the mass function $\mathcal{M}$ defined as:
\begin{equation}\label{mr}
\mathcal{M}(\bar{r})= \int^{\bar{r}}_0 4\pi \tilde{r}^2\frac{\bar{Q}(\tilde{r})}{\bar{\sigma}(\tilde{r})} d\tilde{r},
\end{equation}
appearing in the solution of the static spherical-symmetric metric:
\begin{equation}\label{mod_geo}
 \bar{a}(\bar{r})=\left( 1-\frac{2G \mathcal{M}(\bar{r})}{c^2\bar{r}} \right)^{-1}.
\end{equation}
The prime $'$ in the equation (\ref{tov}) denotes the derivative with respect to the radial coordinate $\bar{r}$.
The quantities $\bar{Q}$ and $\bar{\Pi}$ appearing in the above TOV and mass equations are called effective energy density and pressure, respectively, and are defined as:
\begin{eqnarray}\label{def}
\bar{Q}(\bar{r}):=\bar{\rho}(\bar{r})-\frac{\bar{\sigma}(\bar{r})\bar{W}_{tt}(\bar{r})}{\kappa^2c^2 \bar{b}(\bar{r})},\\
\label{def2} \bar{\Pi}(\bar{r}):=\bar{p}(\bar{r})-\frac{\bar{\sigma}(\bar{r})\bar{W}_{\bar{r}\bar{r}}(\bar{r})}{\kappa^2 \bar{a}(\bar{r})}.
\end{eqnarray}
Here, the energy density $\bar{\rho}$ and pressure $\bar{p}$ are the ones related by the barotropic equation of state, $\bar{p}=\bar{p}(\bar{\rho})$, and appear in the perfect fluid form of the energy momentum tensor $\bar{T}_{\mu\nu}$:
\begin{equation}
    \bar{T}_{\mu\nu}=(\bar{\rho}+\bar{p})\bar{u}_\mu \bar{u}_\nu +\bar{p} \bar{h}_{\mu\nu},
\end{equation}
where $\bar{u}^\mu$ is a vector field co-moving with the fluid while $\bar{h}_{\mu\nu}=\bar{g}_{\mu\nu}+\bar{u}_\mu \bar{u}_\nu$ is a projector tensor on the $3$-dimensional hypersurface.

\subsection{Tolman-Oppenheimer-Volkoff equation}

Using the formalism briefly discussed above, we will write down the components of the $\hat{W}_{\mu\nu}$ tensor considered in the Einstein frame with respect to the metric $\hat g_{\mu\nu}$, which we will use to construct the TOV equations. Its form in the general frame can be found in the Appendix (\ref{appen}). On the other hand, in the Einstein frame this tensor has a much simpler form written with respect to the  scalar-tensor invariants as:
\begin{subequations}
\begin{align}
\hat{W}_{tt} & = \frac{1}{2}\frac{\hat{b}}{\hat{a}}\left(\partial_{\hr} \mathcal{I}\right)^2 + \frac{\hat{b}}{2}\mathcal{I}_2, \\
\hat{W}_{\hr\hr} & = \frac{1}{2}\left(\partial_{\hr} \mathcal{I}\right)^2 - \frac{\hat{a}}{2}\mathcal{I}_2, \\
\hat{W}_{\theta\theta} & = -\frac{1}{2}\frac{\hr^2}{\hat{a}}\left(\partial_{\hr} \mathcal{I}\right)^2 - \frac{\hr^2}{2}\mathcal{I}_2, \\
\hat{W}_{\Phi\Phi}  &= \sin^2{\theta} \hat{W}_{\theta\theta}
\end{align}
\end{subequations}
which now can be used to construct the effective energy density and pressure given in the ST invariants' terms as:
\begin{subequations}
\begin{align}
& \hat{Q}  = \hat{\rho} - \frac{\left(\partial_{\hr} \mathcal{I}\right)^2}{2\kappa^2c^2\hat{a}} - \frac{\mathcal{I}_2}{2\kappa^2c^2}, \\
& \hat{\Pi} = \hat{p} -  \frac{1}{2\kappa^2\hat{a}}\left(\partial_{\hr} \mathcal{I}\right)^2 + \frac{1}{2\kappa^2}\mathcal{I}_2.\label{efP}
\end{align}
\end{subequations}
Let us  remind that $\hat\rho$ and $\hat p$ are not considered as physical quantities. 

Consequently, we are now able to write down the TOV equation in Einstein frame using the schematic way (\ref{tov}):
\begin{equation}
\begin{split}
\frac{d\hat{\Pi}}{d\hat{r}}= & -\frac{G\mathcal{M}}{c^2\hr^2}\left(c^2\hat{\rho}+\hat{p} -\frac{(\partial_r\mathcal{I})^2}{\kappa^2\hat{a}}\right) \left(1+\frac{4\pi \hr^3\hat{\Pi}}{c^2\mathcal{M}}\right)\\ &\times\left(1-\frac{2G\mathcal{M}}{c^2\hr}\right)^{-1} + \frac{1}{\kappa^2 \hat{a}\hr}\left(\partial_{\hr} \mathcal{I}\right)^2,
\end{split}
\end{equation}

\noindent where the effective mass is defined as:
\begin{equation}
\mathcal{M}(\hr)= \int^{\hr}_0 4\pi r^2 \hat{Q}(r) dr.
\end{equation}
In order to rewrite it with respect to the Einstein frame pressure $\hat{p}$ and density $\hat\rho$, we use the definition (\ref{efP}) to get:

\begin{align}\label{tovE}
   \frac{d\hat{p}}{d\hat{r}}=&-\frac{G\mathcal{M}(\hat{r})}{c^2\hr^2}(c^2\hat\rho +\hat p)\left( 1-\frac{2G\mathcal{M}(\hat{r})}{c^2\hr} \right)^{-1}\\
   &\times\left[ 1+\frac{4\pi \hat r^3}{c^2\mathcal{M}(\hat{r})}\left(\hat p-\frac{(\partial_{\hat r}\mathcal{I})^2}{2\kappa^2\hat a}+\frac{\mathcal{I}_2}{2\kappa^2} \right) \right]\nonumber\\
   &+\hat T\partial_{\hat{r}} \ln\mathcal{I}_1-\frac{(\partial_{\hat{r}} \mathcal{I})^2}{\kappa^2 \hat{a}\hat{r}}
\end{align}

\noindent while the effective mass can be expressed as:
\begin{equation}\label{masE}
    \mathcal{M}(\hr)= M_0(\hr)-\eta(\hr,\mathcal{I},\mathcal{I}_2).
\end{equation}
We have defined the "General Relativity" mass in the Einstein frame as $\mathcal{M}_0(\hr)= \int^{\hr}_0 4\pi x^2 \hat{\rho}(x) dx$ and the additional ingredient $\eta(\hr,\mathcal{I},\mathcal{I}_2)$ as:
\begin{equation}
    \eta(\hr,\mathcal{I},\mathcal{I}_2)=\frac{c^2}{4G}\int^{\hr}_0  x^2 \left(\frac{(\partial_{x}\mathcal{I})^2}{\hat a}+\mathcal{I}_2\right)dx.
\end{equation}

The equations (\ref{tovE}) and (\ref{masE}) are TOV equations written in Einstein frame. In order to consider a physical stellar system, we need to transform them to the (physical) Jordan frame, that is, we need
to express the TOV equations in terms of physical metric components and quantities. Thus, they are defined as follows (w.r.t. the initial Wagoner parametrization):
\begin{subequations}
    \begin{align}
        & g_{\mu\nu} = e^{2\bar{\alpha}}\bg_{\mu\nu}, \quad r = e^{\bar{\alpha}}\bar{r}, \\
        & a = (1 - r \partial_r \bar{\alpha})^2 \bar{a}, \quad b = e^{\bar{\alpha}}\bar{b}, \\
        & \rho = \frac{\bar{\rho}}{e^{4\bar{\alpha}}}, \quad p = \frac{\bar{p}}{e^{4\bar{\alpha}}}, \quad T = \frac{\bar{T}}{e^{4\bar{\alpha}}},\label{conftrans}
    \end{align}
\end{subequations}
while the relations between the invariant Einstein and Jordan frame quantities are now:
\begin{subequations}
\begin{align}
    & \hat{r}=\sqrt{\mathcal{I}_1} r, \quad \hat{a} = \mathcal{I}_1 a \left(\frac{dr}{d\hr}\right)^2, \\
    & \hat{\rho}=\frac{\rho}{\mathcal{I}^2_1}, \quad \hat{p}=\frac{p}{\mathcal{I}^2_1}, \quad \hat{T}=\frac{T}{\mathcal{I}^2_1}.\label{conftrans2}
\end{align}
\end{subequations}
Therefore, the effective mass written as a function of the coordinate $r$ and Jordan frame functions $\rho$, $p$ will be: 
\begin{equation}\label{masa}
    \begin{split}
        \mathcal{M}(r) = &\int^r_0  \frac{4\pi r^2}{\sqrt{\mathcal{ I}}_1}\Bigg[{\rho} - \frac{1}{2\kappa^2c^2}\frac{\mathcal{I}_1}{{a}}\left(\partial_r \mathcal{I}\right)^2 - \frac{1}{2\kappa^2c^2}\mathcal{I}^2_1\mathcal{I}_2\Bigg]\\
        &\times\left(\frac{r}{2}\partial_{r} \ln\mathcal{I}_1 + 1 \right)dr,
    \end{split}
\end{equation}
while the TOV equation can be rewritten as: 
\begin{equation}\label{tovJ}
    \begin{split}
        & \frac{dp}{dr} =  \Bigg[-\frac{G\mathcal{M}(r)}{c^2r^2\mathcal{I}^{1/2}_1}(c^2\rho+p)\left(1 - \frac{2G\mathcal{M}(r)}{c^2r\mathcal{I}^{1/2}_1}\right)^{-1}\\
        &\times\left(1 + \frac{4\pi\mathcal{I}_1^\frac{3}{2}r^3}{c^2\mathcal{M}(r)}\left(\frac{p}{\mathcal{I}^2_1} - \frac{\left(\partial_r \mathcal{I}\right)^2 }{2\kappa^2 a \mathcal{I}_1}+ \frac{\mathcal{I}_2}{2\kappa^2}\right)\right) \\
        & -\frac{\mathcal{I}_1^{2}(\partial_r \mathcal{I})^2}{\kappa^2 a r}\Bigg]\left(\frac{r}{2}\partial_{r} \ln\mathcal{I}_1 + 1 \right)
   +\left(-c^2\rho + 5p\right)\partial_{r} \ln\mathcal{I}_1
    \end{split}
\end{equation}

The equations (\ref{masa}) and (\ref{tovJ}) are mass and relativistic hydrostatic equilibrium equations, respectively, describing a spherically-symmetric stellar system in the scalar-tensor theory written with respect to invariants' formalism given by \cite{tartu}.
We recover the GR form for the above equations when the invariant $\mathcal{I}_1=1$ and the remaining ones are set to zeros (or $\mathcal{I}$ is a constant).

\subsection{Non-relativistic limit of TOV equation}\label{non-relat}

Let us consider now non-relativistic limit of the equations (\ref{masa}) and (\ref{tovJ}). Such equations are used to study, for instance, main-sequence stars,  or low-mass stars and substellar objects.
In the non-relativistic limit one assumes that the pressure is much smaller then the energy density
$p \ll c^2\rho$, as well as that the gravitational pressure can be neglected, $r^3p \ll c^2M_0$. The curvature effects are also neglected, that is, $2GM_0/c^2r \ll 1$ holds.

Since we want to keep our equations as general as possible, we will not assume any limiting form of the invariants $\mathcal{I}$ and $\mathcal{I}_2$ when:
\begin{equation}
 \phi(r)\approx \phi_0+\bar{\epsilon} \varphi(r)   
\end{equation}
 is applied to the scalar field in the Jordan frame. However, in order not to change significantly the physics of the Sun, we cannot allow the coupling, which is represented by the invariant $\mathcal{I}_1$, to be too strong in our case. Because of that fact, one then requires that (we assume that $\varphi(0)=0$ and $i_1(0)=0$):
\begin{align}\label{inv1}
    \mathcal{I}_1\approx 1+\epsilon i_1(r).
\end{align}
such that the mass function (\ref{masa}) will have the form $\mathcal{M}=M_0(r)+\mu(r,i_1,\mathcal{I},\mathcal{I}_2)$. Thus, applying (\ref{inv1}) to (\ref{masa}) and keeping only the terms up to the first order, we may write:
\begin{align}\label{mass1}
    &\mathcal{M}(r)=M_0(r)+\epsilon\int^r_0\pi r^2\rho  \Big(  2(r i_1'-i_1)\\
    &\quad\quad\quad- \frac{c^2 \mathcal{I}_2}{8\pi G\rho}(3i_1+ri'_1)    \Big)dr
\end{align}

\noindent where:
\begin{equation}\label{mass2}
    M_0(r)=\int^r_0 4\pi r^2\rho \left[1- \frac{c^2\mathcal{I}_2}{16\pi G\rho}\right]dr.
\end{equation}

\noindent In the above expressions we have
treated the term $(d\mathcal{I}/dr)^2$ as a term of the second order.

Then, in the similar way, the non-relativistic limit of the TOV equations is given by:
\begin{align}
    \frac{dp}{dr} =  -\frac{G\mathcal{M}(r)}{c^2r^2}\rho 
    \left[  1+\epsilon \left(\frac{c^2i_1' r^2}{GM_0}+i'r-\frac{i_1}{2}\right)\right],
\end{align}
where we have skipped the terms such as $\alpha\mathcal{I}_2/M_0c^2$ etc. However, it may happen, similarly as with the terms $(d\mathcal{I}/dr)^2$, that they will have a non-negligible contribution but this depends on the relations between the invariants and matter fields (thus on a particular theory under consideration), given by the Klein-Gordon equation.


\section{Examples}\label{examples}
We will briefly discuss two models of gravity in both, metric and Palatini approaches. Therefore, as a first example we will consider $f(R)$ gravity and then Brans-Dicke model.
\subsection{$f(R)$ gravity}

Let us consider as an example the $f(R)$ theory in order to illustrate how the invariant quantities allow one to quantify the difference between theories of gravity. In this section, we will compute the invariants for both metric and Palatini versions of the theory. 

First, let us notice that $f(R)$ gravity has a scalar-tensor representation if $\frac{d^2 f}{dR^2}\neq 0$. Assuming this condition is satisfied, we can perform a Legendre transformation and identify a scalar degree of freedom with the derivative of the $f$ function w.r.t the curvature $R$. This procedure is canonical and has been presented in a number of papers, for example in \cite{capo2011}. The result is the following:

\begin{equation}
\begin{split}
S[g_{\mu\nu},  \Phi, \chi] = & \frac{1}{2\kappa}\int_\Omega d^4x \sqrt{-g}\Big[\Phi R(g) +\delta_\Gamma\frac{3}{2\Phi}g^{\mu\nu}\partial_\mu\Phi\partial_\nu\Phi\\
&- V(\Phi)\Big] + S_\text{matter}[g_{\mu\nu}, \chi].
\end{split}
\end{equation}

\noindent Here, the parameter $\delta_\Gamma$ assumes two possible values:
\begin{equation}
\delta_\Gamma = 
\begin{cases}
1 \text{   for the Palatini theory}\\
0 \text{   for the metric theory.}
\end{cases}
\end{equation}

\noindent We can immediately identify the functions of the scalar field defining the theory:
\begin{align}
\mathcal{A}(\Phi) &=\Phi, \quad \mathcal{B}(\Phi) = -\delta_\Gamma\frac{3}{2\Phi}, \nonumber\\ \mathcal{V}(\Phi) &= V(\Phi), \quad \alpha(\Phi) = 0.
\end{align}

The invariants $\mathcal{I}_1$ and $\mathcal{I}_2$ will be the same for both approaches; however, we will be able to distinguish between the two formalism by looking at the values of the invariants $\mathcal{I}$ and $\tilde{\mathcal{I}}$. Their values are given in the table below:

\begin{table}[H]
\caption{Invariant scalar field for $f(R)$ gravity in different approaches.}
\centering
\begin{tabular}{lll}
\hline\noalign{\smallskip}
Invariant & \text{Metric} & \text{Palatini} \\
\noalign{\smallskip}\hline\noalign{\smallskip}
$\mathcal{I}$ & $\sqrt{\frac{3}{4}}\ln \left(\frac{\Phi}{\Phi_0}\right)$ & $0$ \\
$\tilde{\mathcal{I}}$ & $0$ & $\sqrt{6}(\sqrt{\Phi} - \sqrt{\Phi_0})$ \\
\noalign{\smallskip}\hline
\end{tabular}
\end{table}

We can be more specific and take as an example the following model: $f(R) = R + \beta R^n$, where $n\in \mathbb{R}$, $n\neq 1$. Such a choice will lead to the following parametrization in the ST representation:

\bgroup
\def\arraystretch{1.3}
\begin{table}[h]
\centering
\begin{tabular}{c c c c c}
\hline
 &  $\ca$ & $\cb$ & $\cv$ & $\alpha$ \\
\hline
\text{metric} & $\Phi$ & $0$ & $\left(n^{-\frac{1}{n-1}} - n^{-\frac{n}{n-1}}\right)\frac{(\Phi - 1)^\frac{n}{n-1}}{\beta^{\frac{1}{n-1}}}$ & $0$ \\
\text{Palatini}  & $\Phi$ & $-\frac{3}{2\Phi}$ & $\left(n^{-\frac{1}{n-1}} - n^{-\frac{n}{n-1}}\right)\frac{(\Phi - 1)^\frac{n}{n-1}}{\beta^{\frac{1}{n-1}}}$ & $0$ \\
\hline
\end{tabular}
\caption{\small{Parametrization of the $f(R) = R + \beta R^n$ model.}}
\end{table}

\noindent and, consequently, to the invariants:

\bsub
\begin{align}
& \mathcal{I}_1(\Phi) = \Phi, \\
& \mathcal{I}_2(\Phi) = \beta^{-\frac{1}{n-1}}\left(n^{-\frac{1}{n-1}} - n^{-\frac{n}{n-1}}\right)\frac{(\Phi - 1)^\frac{n}{n-1}}{\Phi^2}.
\end{align}
\esub

For the Palatini theory, scalar field is non-dynamical, so the invariant $\hi$ cannot be used in the theory. Therefore, the relation between the field and the trace of energy-momentum tensor will be purely algebraic, and the exact dependence can be obtained from the equation (\ref{invEinEOM2}), with the derivative taken w.r.t. the scalar field $\Phi$. All the remaining invariants entering the field equations and, consequently, the TOV equation, will depend explicitly on $\hat{T}$. 

In the Palatini case, the effective energy density and pressure will have a simple form:
\bsub
\begin{align}
\hat{Q}_P & = \hat{\rho} - \frac{1}{2\kappa^2c^2}\mathcal{I}_2(\hat{T}), \\
\hat{\Pi}_P & = \hat{p} + \frac{1}{2\kappa^2}\mathcal{I}_2(\hat{T}).
\end{align}
\esub

\noindent The TOV equation in the Einstein frame will now read as:
\begin{equation}
\begin{split}
\frac{d\hat{\Pi}_P}{d\hat{r}}= & -\frac{G\mathcal{M}_P}{c^2\hr^2}\left(c^2\hat{\rho}+\hat{p}\right) \left(1+\frac{4\pi \hr^3\hat{\Pi}_P}{c^2\mathcal{M}_P}\right)\left(1-\frac{2G\mathcal{M}_P}{c^2\hr}\right)^{-1},
\end{split}
\end{equation}
which agrees with \cite{awstab}.
If we set $n=2$, we recover Starobinsky theory. For this particular choice, the invariants $\mathcal{I}_1$ and $\mathcal{I}_2$ can be expressed in terms of the trace of energy-momentum tensor in the Jordan frame: 
\bsub
\begin{align}
    & \mathcal{I}_1 = \Phi = 1 + 4\beta \kappa^2 (c^2\rho - 3p), \\
    & \mathcal{I}_2 = \frac{4\beta\kappa^4(c^2\rho - 3p)^2}{\left(1 + 4\beta \kappa^2 (\rho - 3p) \right)^2}.
\end{align}
\esub

\noindent The mass function will take the form: 
\begin{equation}
\begin{split}
    \mathcal{M}_P(r) = &   \int^{r}_0 4\pi\tilde{r}^2 \frac{\rho - 2c^{-2}\beta\kappa^2(c^2\rho - 3p)^2}{\left(1 + 4\beta \kappa^2 (c^2\rho - 3p) \right)^{1/2}}\\
    &\times\left[1+ \frac{\tilde{r}}{2}\partial_{\tilde{r}}\ln\left(1+4\beta\kappa^2 (c^2\rho -3p)\right)\right]d\tilde{r}
\end{split}
\end{equation}
In the non-relativistic limit, the TOV equation (\ref{tovJ}) for quadratic Palatini gravity takes the form
\begin{equation}\label{tovpal}
     \frac{dp}{dr} = -\frac{G\mathcal{M}_P(r)}{r^2}\rho 
    \left[  1+4c^2\beta\kappa^2 \left(r\rho'-\frac{\rho}{2}+\frac{c^2r^2\rho'}{GM_0}\right)\right].
\end{equation}
whereas the mass reduces to:
\begin{equation}
     \mathcal{M}_P(r) =    \int^{r}_0 4\pi\tilde{r}^2\rho\left[1+2c^2\kappa^2\beta (-2\rho + \tilde{r} \rho')\right]d\tilde{r}
\end{equation}

The expression (\ref{tovpal}) has a slightly different form (in the coefficients in the round bracket) than the result obtained in \cite{Wojnar:2018hal}. This is because of the fact that in \cite{Wojnar:2018hal}, the assumption of the invariant polytropic equation of state for the quadratic Palatini model was made, following the result given in \cite{mana}. In this work we do not set any particular relation between pressure and density, neither we assume its behaviour under the conformal transformation.

In the metric case, the relation between the scalar field and the invariant $\hi$ becomes relevant. It must be used to express the invariants $\mathcal{I}_1$ and $\mathcal{I}_2$ as a function of $\hi$. The result is the following:
\begin{subequations}
\begin{align}
& \mathcal{I}_1(\hi) = \Phi_0 e^{\frac{2}{\sqrt{3}}\hi},\\
& \mathcal{I}_2(\hi) = \beta^{-\frac{1}{n-1}}\left(n^{-\frac{1}{n-1}} - n^{-\frac{n}{n-1}}\right)\frac{(\Phi_0 e^{\frac{2}{\sqrt{3}}\hi} - 1)^\frac{n}{n-1}}{\Phi_0^2 e^{\frac{4}{\sqrt{3}}\hi}}.
\end{align}
\end{subequations}

Unlike in the Palatini case, it is now not possible to express $\mathcal{I}$ as an algebraic function of $\rho$ and $p$, as the relation between these quantities is described by the Klein-Gordon equation with a source. Applying the above forms to the equations (\ref{masa}) and (\ref{tovJ}), one obtains the mass function and TOV equation for the $f(R)$ metric gravity. Let us notice that the equation (\ref{tovJ}) already includes the Klein-Gordon equation to get rid of the second derivative of $\mathcal{I}$. However, we do not write the TOV equations explicitly because of their complex form in this particular case.

It was shown in \cite{Capo2020}-\cite{Capo2021} that for
$f(R)$ gravity, one can use multimessenger astronomy results to demonstrate the maximal mass and causal limit maximal mass of neutron stars, for example in Starobinsky model. As the analysis is usually performed in the Jordan frame, the use of invariants might help to generalize the investigation by allowing one to express relevant quantities in a frame-independent way, and transform solutions easily between frames.

\subsection{Brans-Dicke theory}

Let us now consider a scalar-tensor theory which cannot be derived from $f(R)$ gravity, namely, a Brans-Dicke gravity without the self-interaction potential, both in the metric and the Palatini approach. The action for the theory is the following:

\begin{equation}\label{BDaction}
\begin{split}
S[g, \Gamma, \Phi] & = \frac{1}{2\kappa^2}\int_\Omega d^4x\sqrt{-g}\left[\xi\Phi R(g, \Gamma) - \frac{\omega}{\Phi}g^{\mu\nu}\partial_\mu\Phi\partial_\nu\Phi\right]\\
&+ S_\text{matter}[g, \chi].
\end{split}
\end{equation}

Here, $\xi$ and $\omega$ are parameters of the theory. We assume that $\xi > 0$ and $\omega \neq 0$, otherwise the scalar field would have no dynamics in the Palatini formulation. Also, the action (\ref{BDaction}) is not specified to be either a Palatini, or a metric action; therefore, we take a theory with the same functions of the scalar field, but with different definitions of the curvature scalar. 

In the Palatini case, one needs to consider an additional variation with respect to the independent conection. As a result of this procedure, one discovers that the initially independent connection is in fact Levi-Civita with respect to a new metric tensor, conformally related to the initial one. The connection introduces no additional degrees of freedom, and can be eliminated from the action, modifying the kinetic coupling of the scalar field, $\cb$, yielding effectively a metric theory. The scalar field functions in this case can be written as:

\begin{equation}
\ca = \xi \Phi, \quad \cb = \frac{\omega}{\Phi} - \delta_\Gamma\frac{3\xi}{2\Phi},\quad \cv=0,\quad \alpha = 0,
\end{equation}

\noindent where $\delta_\Gamma$ was defined in the previous section. The invariants are now:
\bsub
\begin{align}
&\mathcal{I}_1(\Phi) = \xi\Phi, \\
&\mathcal{I}_2(\Phi) = 0, \\
&\mathcal{I} = \sqrt{\frac{2\omega + 3\xi(1 - \delta_\Gamma)}{4\xi}}\ln\left(\frac{\Phi}{\Phi_0.}\right)
\end{align}
\esub

How the scalar field varies will be now described by a massless Klein-Gordon equation with a source term:

\begin{equation}
\hat{\Box}\mathcal{I} = \kappa^2\sqrt{\frac{4\xi}{2\omega + 3\xi(1 - \delta_\Gamma)}}\hat{T}.
\end{equation}

As we can see, the invariant scalar field is sourced by the trace of energy-momentum tensor. In case of radiation, the trace vanishes, and the equation is of particularly simple form. 

The components of the $\hat{W}_{\mu\nu}$ tensor for this class of theories are now:
\begin{subequations}
\begin{align}
\hat{W}_{tt} & = \frac{1}{2}\frac{\hat{b}}{\hat{a}}\left(\partial_{\hr} \mathcal{I}\right)^2, \\
\hat{W}_{\hr\hr} & = \frac{1}{2}\left(\partial_{\hr} \mathcal{I}\right)^2, \\
\hat{W}_{\theta\theta} & = -\frac{1}{2}\frac{\hr^2}{\hat{a}}\left(\partial_{\hr} \mathcal{I}\right)^2, \\
\hat{W}_{\Phi\Phi}  &= \sin^2{\theta} \hat{W}_{\theta\theta},
\end{align}
\end{subequations}
and the effective energy density and pressure can be expressed as: 
\begin{subequations}
\begin{align}
& \hat{Q}  = \hat{\rho} - \frac{1}{2\kappa^2c^2\hat{a}}\left(\partial_{\hr} \mathcal{I}\right)^2, \\
& \hat{\Pi} = \hat{p} -  \frac{1}{2\kappa^2\hat{a}}\left(\partial_{\hr} \mathcal{I}\right)^2,
\end{align}
\end{subequations}

where $\hat{\rho} = \frac{\rho}{\xi^2\Phi^2}$, and $\hat{p} = \frac{p}{\xi^2\Phi^2}$.
As for the metric case, we do not write the hydrostatic equilibrium equation explicitly because of its long form.

\section{Discussion and conclusions}

In the presented paper we have obtained the stellar structure equations for a general family of the modified gravity theories, written with respect to the scalar field invariants, introduced some time ago in \cite{tartu}. We found out the invariant representation very useful since it allows to study diverse theories of gravity, such as for example $f(R)$ and scalar-tensor gravity in different approaches - metric, Palatini, or hybrid, at one go \cite{borow2020}.
The relativistic hydrostatic equilibrium equation (\ref{tovJ}) and the mass function (\ref{masa}) are thus the master equations of our work, and their specific shape can help us to understand effects arising by the presence of such invariants, whose specific forms and characteristics, as shown by a few examples in the section \ref{examples}, are ruled by gravitational theories, which possess the invariant representation.

Therefore, one may also try to interpret the role of the invariants, as it was presented for example in \cite{Velten:2016bdk}, where the free parametrization of the TOV equation was discussed together with physical meanings of the parameters. Let us firstly notice that the invariant $\mathcal{I}$, which carries an information about the varying scalar field, and $\mathcal{I}_2$, which traces its potential, enter the part of the stellar equation (\ref{tovJ}) related to the active gravitational effects of pressure (studied for instance in \cite{schwab}). In the case of GR, this pressure is a pure relativistic effect which does not appear in the non-relativistic counterpart of the hydrostatic equilibrium equation. Thus, modifications which can be related to the varying scalar field and/or its potential, reinforce or impair, depending on their nature, the effect of the gravitational pressure. Let us also comment that the invariant $\mathcal{I}_2$ also appears in the TOV equation (\ref{tovJ}) throughout the mass function. We will discuss it in a moment.

On the other hand, the non-minimal coupling, represented by the invariant $\mathcal{I}_1$, clearly plays many roles in the stellar structure. It also appears in the gravitational pressure part, in a non-trivial way, strengthening or weakening this relativistic effect. It acts, as already mentioned in the introduction, as an effective gravitational constant which is also apparent in our TOV equation. Moreover, it contributes to the inertial pressure and intrinsic curvature contributions which are, similar to the gravitational pressure, relativistic impacts, affecting equation of state (see such an effect in some theories of gravity, for example \cite{kim}-\cite{Wojnar:2020txr}) and geometry, where the last one is an expected result. Besides, the presence of $\mathcal{I}_1$ may also be interpreted as anisotropies of the fluid (pronounced by the last terms of (\ref{tovJ})) 
\cite{Afonso:2018bpv,Afonso:2018hyj}. 

Let us notice that the anisotropy of the fluid, as well as the active gravitational effects of pressure provided by the presence of the invariants $\mathcal{I}$ and $\mathcal{I}_1$ arise in the Einstein frame, too. 

The mass equation (\ref{masa}) also provides some information about the nature of the invariants. The most prominent one is the fact that $\mathcal{I}_1$ cannot be a strong coupling; that is, it cannot differ significantly from unity in order not to violate the solar physics. It may happen, as it was discussed in, for instance, \cite{asta1}-\cite{review}, that the extra terms introduced by the invariants provide a non-zero contribution to the mass function far away from the star's surface (when $p(R)=0$, where $R$ is the star's radius); that is, the mass function does not converge when $r\rightarrow\infty$ but can oscillate instead around a constant value. This fact has following consequences: as proposed in \cite{asta1} and then discussed in more detail in \cite{oliver}, in the case of $f(R)$ metric gravity and general ST theories which introduce an additional degree of freedom, this phenomena could be used to test such theories against GR by the mean of the surface gravitational redshift. It comes from the fact that mass measured by a distant observer (gravitational mass) differs from the stellar mass, which is understood as a sphere bounded by the star's surface together with gravitational energy, in the energy of the extra degree of freedom, which happens not to be zero inside and outside the star. Moreover, it was also demonstrated that the stellar mass appearing in the definition of the mentioned redshift in the case of modified gravity is smaller that its analogue in GR. Therefore, a high-precision in the measurements of the surface redshift can be in future a powerful tool to test and to discriminate different theories of gravity.

On the other hand, the invariant $\mathcal{I}_2$ in the masses (\ref{mass1}) and (\ref{mass2}), as well as in the TOV equation (\ref{tovJ}), can be identified with the cosmological constant \cite{liu,kain};  then, its contribution is negligible to the solar gravitational mass $\mathcal{M}$ while the input from $i_1$ is also expected to be small, as already mentioned.

It should be also highlighted that $i_1$ is not an invariant anymore. As in the case of the invariants, we do not know its form without assuming any particular theory (in that case, the potential of the scalar field), or without guessing/adjusting its character by other means. Therefore, 
in order to write down the non-relativistic hydrostatic equilibrium equation in the terms of the matter fields only, one should choose a particular model of gravity. Then, the modified Klein-Gordon equation will provide the relation between the invariants (the scalar field) and the trace of the energy momentum tensor, that is, the matter fields. Equipped with this relation, it will be possible to write hydrostatic equilibrium equation in such a form that it would allow to study physical stellar system. However, as briefly discussed in the example section, it might be quite tough because of the complicated form of the equations.

Since the theories discussed in this work introduce an additional scalar field permeating the space-time, it will be necessary to invoke a screening mechanism effectively hiding its effect outside stars. As it was shown in recent papers, the mechanism can be broken inside stellar objects in more general scalar-tensor theories, such us DHOST or beyond Horndeski, where the gravitational potential is not determined solely by the enclosed mass, but depends additionally on extra terms \cite{lan},\cite{koy}-\cite{koyama}.
Moreover, another issue concerns Noether symmetries in scalar-tensor theories, discussed for example in \cite{Capo2007}-\cite{MDL2018} in the context of spherically-symmetric objects. In the most general case, the invariant scalar field carries an additional degree of freedom that will have a non-zero contribution to Noether conserved quantities. Because of the very nature of invariants, all solutions would be relevant for whole conformally-equivalent classes of theories.
We did not investigate these issues in this paper. Work along this line is currently underway.

\section*{Acknowledgements}
   The authors would like to thank Andrzej Borowiec, Laur J\"{a}rv, and Margus Saal for useful comments. We are also grateful to an anonymous referee for suggestions concerning future research. 
	
    AK is a beneficiary of the Dora Plus Program, organized by the Unversity of Tartu.
	AW is supported by  the  EU  through  the  European  Regional  Development  Fund  CoE  program  TK133  "The  Dark  Side of the Universe".

\appendix
\section{Components of the tensor $W_{\mu\nu}$ in the general frame}\label{appen}
For the general case of metric scalar-tensor gravity, the tensor $\bar{W}_{\mu\nu}$ takes the following form:

\begin{equation}
\begin{split}
& \bar{W}_{\mu\nu} = -\frac{1}{\baa}\left(\frac{1}{2}\bb + \baa''\right)\bg_{\mu\nu}\bg^{\alpha\beta}\partial_\alpha\bp\partial_\beta\bp\\
&+ \frac{1}{\baa}\left(\bb + \baa''\right)\partial_\mu\bp\partial_\nu\bp + \frac{\baa'}{\baa}(\bg_{\mu\nu}\bar{\Box} - \bar{\nabla}_\mu\bar{\nabla}_\nu)\bp \\
& -\frac{1}{2}\frac{\bar{\mathcal{V}}}{\baa}\bg_{\mu\nu},
\end{split}
\end{equation}
and the $\sigma$ function can be identified with $\baa$.

For the spherical-symmetric object 
\begin{equation}\label{metric}
 ds^2=-\bar{b}(\bar{r})dt^2+\bar{a}(\br)d\br^2+\br^2d\theta^2+\br^2\sin^2{\theta} d\phi^2.
\end{equation}
the relevant components of the $\bar{W}_{\mu\nu}$ tensor are given by:

\begin{subequations}
\begin{align}
\begin{split}
\bar{W}_{tt} = & \frac{1}{\baa}\left(\frac{1}{2}\bb + \baa''\right)\frac{\bar{b}}{\bar{a}}\left(\frac{d\bp}{d\br}\right)^2 \\
&+ \frac{\baa'}{\baa}\Bigg(\sqrt{\frac{\bar{b}}{\bar{a}}}\frac{d}{d\br}\sqrt{\frac{\bar{b}}{\bar{a}}}+ \frac{2}{\br\:\bar{a}} - \frac{1}{2\bar{a}}\frac{d\bar{b}}{d\br}\Bigg)\frac{d\bp}{d\br}\\ 
&+  \frac{\baa'}{\baa}\frac{\bar{b}}{\bar{a}}\frac{d^2\bp}{d\br^2} + \frac{1}{2}\frac{\bv}{\baa}\bar{b},
\end{split} \\
\begin{split}
\bar{W}_{\br\br} = & \frac{\bb}{2\baa}\left(\frac{d\bp}{d\br}\right)^2 - \frac{\baa'}{\baa}\Bigg(\sqrt{\frac{\bar{a}}{\bar{b}}}\frac{d}{d\br}\sqrt{\frac{\bar{b}}{\bar{a}}}+ \frac{2}{\br} + \frac{1}{2\bar{a}}\frac{d\bar{a}}{d\br}\Bigg)\frac{d\bp}{d\br}\\
&-\frac{1}{2}\frac{\bv}{\baa}\bar{a},
\end{split} \\
\begin{split}
\bar{W}_{\theta\theta} = & -\frac{1}{\baa}\left(\frac{1}{2}\bb + \baa''\right)\frac{\br^2}{\bar{a}}\left(\frac{d\bp}{d\br}\right)^2\\
&- \frac{\baa'}{\baa}\Bigg(\br^2\sqrt{\frac{1}{\bar{a}\:\bar{b}}}\frac{d}{d\br}\sqrt{\frac{\bar{b}}{\bar{a}}}+ \frac{\br}{\bar{a}}\Bigg)\frac{d\bp}{d\br}\\
&-  \frac{\baa'}{\baa}\frac{\br^2}{\bar{a}}\frac{d^2\bp}{d\br^2}- \frac{1}{2}\frac{\bv}{\baa}\br^2,
\end{split} \\
\begin{split}
\bar{W}_{\Phi\Phi} = & \sin^2{\theta} \bar{W}_{\theta\theta}.
\end{split}
\end{align}
\end{subequations}
Thus, we are happy that we do not need to use them in order to write down the TOV equations.


\end{document}